\ifdefined\pdfminorversion\pdfminorversion=7\fi % Match the supplied figure PDFs (PDF 1.7).
\documentclass[10pt]{article}
\usepackage[utf8]{inputenc}
 \usepackage[a4paper,top=0.8in, bottom=0.7in, left=0.7in, right=0.8in]{geometry}
 \usepackage{appendix}
\usepackage{graphicx}
\usepackage{authblk}
\usepackage[numbers,sort&compress]{natbib}
\usepackage{natbib}
\usepackage{pdflscape}
\usepackage{url}
\usepackage{bm}
\usepackage{color}
\usepackage{amssymb}\usepackage{amsmath,amsthm}
\usepackage{amsmath} \usepackage{multirow}
\usepackage{mathrsfs}

\usepackage{array} % 导言区
\usepackage{caption}
\graphicspath{ {./images/} }

\numberwithin{equation}{section} 
\usepackage{amscd,epstopdf}
\usepackage{graphicx}\usepackage{multirow}
\usepackage{float}
\usepackage{subcaption}

\usepackage{afterpage} \usepackage{rotating}
\usepackage{enumerate}
 
 \newtheorem{proposition}{Proposition}[section]

 \newtheorem{definition}{Definition}[section]
 \usepackage{setspace}
\usepackage{indentfirst} 
\usepackage{graphicx}
\usepackage{bibentry}

\begin{document}

\title{Generalized Emergence through Coarse-Graining:\\ A Category-Theoretic Framework}
\author[1]{Kaiwei Liu}
\author[1]{Jiang Zhang \thanks{Email: zhangjiang@bnu.edu.cn}}
\affil[1]{ School of Systems Science, Beijing Normal University, 100875, Beijing, China}
\affil[2]{Swarma Research, 102300, Beijing, China
}
 \maketitle

\begin{abstract}
\noindent
Quantitative theories of emergence use diverse representations and indicators, making their shared structure difficult to identify.
This paper develops a category-theoretic framework for generalized emergence through coarse-graining, without presupposing dynamics, probability, interventions, or specific information measures.
System descriptions are represented by categories and coarse-grainings by functors.
Generalized emergence requires improvement of a specified description attribute along an admissible, nontrivial functor; an extension evaluates corresponding internal morphisms.
The criterion separates structural admissibility from attribute improvement and accommodates several attributes under a chosen comparison order.
Different functors embody different observational perspectives, making emergence observer-dependent.
Selecting an optimal coarse-graining can partially reduce this dependence, but only under a given attribute and constraints; optimality guarantees neither uniqueness nor emergence.
For dynamical models, state and dynamical coarse-graining are defined separately and linked by strict or approximate closure; unification by one functor is a special case.
Computational mechanics, causal emergence, and approximate dynamical reversibility illustrate the dynamical case.
Two static examples show the criterion at work: symmetry emergence, where coarse-graining polygons improves rotational and dihedral symmetry, and cohesion emergence, where coarse-graining networks raises global efficiency, edge density, and clustering.
Each example also contains an equally coarse case that does not qualify: for polygons, one of the two symmetry measures improves while the other worsens, and for networks, all three cohesion attributes decline.

\end{abstract}

\noindent\textbf{Keywords:} generalized emergence; generalized emergence in dynamical systems; category theory; coarse-graining; functors; attribute lifting; levels of description.

\section{Introduction}
Emergence is a cross-disciplinary commonplace \cite{Holland2000,ClaytonDavies2006}.
Phase transitions produce order parameters that no single molecule possesses \cite{Anderson1972};
flocks, schools, and insect colonies display collective motion and a division of labour that no individual exhibits \cite{Reynolds1987,Couzin2003};
neural populations carry representations that single neurons do not;
markets, cities, and ecosystems generate regularities that are invisible in any of their members \cite{Schelling1978,Johnson2001}.
The vocabulary differs from field to field, the pattern does not:
after some form of organization, a description at a higher level exhibits structures, capabilities, or regularities that the lower-level description does not make visible.

What emergence \emph{is}, however, has remained unresolved.
The philosophical literature separates weak from strong emergence and asks whether higher-level properties are in principle derivable from lower-level ones \cite{Lewes1875,Broad1925,Bedau1997,Chalmers2006}.
This concerns what is deducible in principle, which cannot be operationalized as a testable criterion for a given model or dataset \cite{Carroll2024}.
Consequently, whether a given system is emergent has largely remained a matter of interpretation rather than a property of a description that one could decide \cite{Corning2002}.

Over the last decades, a series of quantitative frameworks has tried to supply such a criterion.
Crutchfield's computational mechanics, also called the theory of causal states, summarizes a stochastic process by the equivalence classes of pasts that yield the same predictive distribution; the statistical complexity of these causal states measures how much information about the past the process stores, and emergent structure is read off as a predictive architecture that the raw microscopic histories do not display \cite{CrutchfieldYoung1989,Crutchfield1994}.
Seth's G-emergence uses nonlinear Granger causality and calls a macroscopic variable emergent when it is both caused by the microscopic variables and autonomous with respect to them \cite{seth2008measuring}.
Hoel, Albantakis, and Tononi compare the effective information \cite{Hoel2013} of a causal model before and after coarse-graining and show that the macroscopic model can carry more effective information than the microscopic one, that is, macro can beat micro \cite{Hoel2013,Hoel2017,Klein2020,Liu2024}.
Rosas et al.\ decompose the information in a multivariate system into redundant, unique, and synergistic components, and identify causal emergence with the dominance of synergy \cite{Rosas2020}.
Zhang et al.\ base the comparison on the singular value decomposition of a transition matrix and show that coarse-graining can increase approximate dynamical reversibility and information transmission efficiency \cite{Zhang2025}.

These criteria are not equivalent, and they can rank the same system differently: effective information, statistical complexity, synergy, Granger-causal gain, and a singular-value spectrum are different quantities.
They nevertheless motivate a common structural question: how does a chosen property change when a microscopic description is replaced by a macroscopic one through coarse-graining?
The framework developed here addresses attributes of descriptions and, in an extension, attributes of morphisms inside those descriptions; it does not claim to cover every criterion in the broader emergence literature.
The individual theories leave one question open:
is there a common formal core behind them, and can it be stated once, without committing in advance to any of their indicators?

Category theory is the natural candidate for such a statement, and categorical treatments of emergence already exist.
Adam formalizes generativity as the failure of an observation functor to preserve the colimits that express interconnection, and quantifies it homologically \cite{Adam2017,AdamDahleh2019}; Li et al.\ turn the resulting cohomological dimension into a computable measure on networks \cite{Li2025}.
A second line, causal abstraction, represents a coarse-graining between two causal models as a natural transformation and asks when the high-level model reproduces the low-level one under interventions \cite{Otsuka2022,Geiger2025,Massidda2023,EnglbergerDhami2025,LorenzTull2026}.
A third characterizes an emergent whole algebraically, as a structured construct together with a functor that forgets its operations \cite{LaGuardia2022}.
These works supply the language we need, but they formalize consistency, compositionality, or soundness rather than improvement.
Consistency and soundness are met by degenerate coarse-grainings, such as collapsing every microscopic state into a single macroscopic one, while non-compositionality is met by harmful ones, such as cascading failure.
None of these criteria therefore makes ``the macroscopic level is better'' the object of the definition.

This paper studies generalized emergence through coarse-graining, taking the shared structure of cross-level attribute comparisons as its object of study.
We represent a description by a category and a coarse-graining by a functor between description categories, and define generalized emergence as an attribute lift along an admissible and nontrivial functor $Q$.
The claim is improvement of a specified property under a specified transformation, not unconditional superiority of the macroscopic description.
The same microscopic description can yield different emergence judgments under different coarse-grainings; emergence is therefore explicitly relative to $Q$ and the attribute $P$.
For dynamical modeling we distinguish state and dynamical coarse-graining, which need not combine into a single functor without additional compatibility.
No probability, intervention, entropy, linear structure, or physical scale is presupposed, so that dynamical theories and static structural comparisons supply applications rather than premises.
The non-dynamical examples include cases with and without attribute lifting, rather than only favorable reductions.
Both polygon cases compare 81 vertices with 9 vertices, and both network cases compare 450 nodes with 9 community nodes, yet their attribute changes differ in direction.
This makes the role of microscopic structure explicit and distinguishes a joint multi-attribute lift from improvement in only one component.

Our main contributions are the following.
\begin{enumerate}
  \item We build the framework itself: a description is a category and a coarse-graining is a functor between description categories, with an optional meta-level that organizes several descriptions; generalized emergence is an attribute lift along an admissible and nontrivial functor, so that structural admissibility and attribute improvement are separate conditions, and the comparison extends from descriptions to the morphisms inside them.
  \item We make the dependence on the coarse-graining explicit: different functors embody different observational perspectives, an optimal coarse-graining chosen by maximizing a given attribute can reduce this dependence only under that attribute and its constraints, and the conditions under which the optimum is unique are stated separately.
  \item We endogenize dynamics inside the same framework: states are generalized elements, state coarse-graining and dynamical coarse-graining are defined separately and linked by strict or approximate closure, and handling both with one functor is a strict special case; computational mechanics, causal emergence, and approximate dynamical reversibility then appear as instances rather than premises.
  \item We test the criterion on two non-dynamical contrasts, symmetry across polygon coarse-grainings and cohesion across network coarse-grainings, each containing one case that satisfies the criterion and one that does not at the same reduction level, which shows that emergence follows from the attribute comparison rather than from reduction as such.
\end{enumerate}
Section~2 develops the framework, Section~3 presents dynamical and non-dynamical instances, and Section~\ref{sec:discussion} discusses the framework's advantages and limitations.
We do not claim that category theory decides which attributes matter.
The framework fixes what levels and relations an emergence judgment requires, and leaves to each scientific community what counts as an attribute, as an admissible coarse-graining, and as a meaningful comparison.

\section{Theories}
\subsection{Scope and Basic Distinctions}

\subsubsection{Systems and Descriptions}
We use ``system'' to denote the object under study, without binding it to a unique mathematical representation.
A system can have multiple descriptions, each preserving only a portion of the system's structure.
Throughout this paper, $\mathcal{S}$ denotes the system itself, and $\mathcal{S}$ is used only as the index under which the descriptions of that system are collected.
A single description of the system is a category $\mathcal{C}$ carrying the basic objects and morphisms.
Any further structure that the research purpose requires is not given a separate symbol:
it enters through the property that an admissible coarse-graining must preserve.
Some dynamics may be represented directly by morphisms in $\mathcal{C}$; for probabilistic dynamics, we choose an appropriate structured category such as a Markov category, a convex category, or another suitable semantic category to determine the corresponding morphisms.

Objects of $\mathcal{C}$ may represent entities, states, state spaces, parts, variables, structural units, or other research objects.
Morphisms may represent relations, transformations, actions, embeddings, compositions, deterministic evolution, or stochastic processes.
This paper does not require that every category carries a dynamical semantics, but it also does not exclude dynamics from within categories: in an appropriate semantic category, dynamics is given by morphisms.

An important limitation is built into this wording:
a category is not the unique ``real ontology'' of the system; it is a structurally organized description of the system.
For the same system, different choices of $\mathcal{C}$ may exist.
Our generalized emergence focuses precisely on the nontrivial transformations between such descriptions and on how attributes change across them.

\subsubsection{Microscopic and Macroscopic Levels as Description Levels}
Let $\mathcal{C}_\mu$ denote a microscopic description of a system and $\mathcal{C}_M$ a macroscopic description.
Subscripts $\mu$ and $M$ primarily indicate levels of description, not physical sizes.
They also do not imply that the macroscopic level necessarily has an independent ontological entity.
Microscopic descriptions may preserve more distinctions; macroscopic descriptions may identify several microscopic objects as one class, ignore some morphisms, or reorganize existing relations.
Therefore, a macroscopic description is not merely ``deleting'' part of the micro objects; it may be a new structured representation.

In the simplest framework, it suffices to provide two categories and a functor from micro to macro:
\begin{equation}\label{functor_micro_macro}
    Q: \mathcal{C}_\mu \to \mathcal{C}_M.
\end{equation}
Here $Q$ acts on both objects(e.g., $A$) and morphisms(e.g., $f,g$) and satisfies
\begin{equation}\label{functor_objects_morphisms}
Q(\mathrm{id}_A)=\mathrm{id}_{Q(A)},\qquad
Q(g\circ f)=Q(g)\circ Q(f).
\end{equation}
Thus, coarse-graining is not an arbitrary mapping of object sets; it is a structural process responsible for composite structure.

\subsubsection{A Description Category as an Optional Meta-Level}
If we only study one microscopic description, one macroscopic description, and a coarse-graining functor, introducing a description category is not necessary.
To unify multiple admissible descriptions of the same system, we may further define an optional description category $\mathrm{Desc}(\mathcal{S})$.
Its objects are the admissible structured descriptions of the system $\mathcal{S}$; each description is a category.
Its morphisms are the structure-preserving conversions allowed between descriptions.

In this meta-level, we treat $\mathcal{C}_\mu$ and $\mathcal{C}_M$ as objects of $\mathrm{Desc}(\mathcal{S})$ and regard $Q$ as a morphism in $\mathrm{Desc}(\mathcal{S})$:
\begin{equation}\label{Desc_S}
\mathcal{C}_\mu, \mathcal{C}_M \in \mathrm{Ob}\bigl(\mathrm{Desc}(\mathcal{S})\bigr),\qquad
Q \in \mathrm{Mor}_{\mathrm{Desc}(\mathcal{S})}(\mathcal{C}_\mu, \mathcal{C}_M).
\end{equation}
There are two distinct ``arrow'' interpretations of $Q$: within descriptions, $Q$ is a functor between categories; at the meta-level, $Q$ is a morphism in the description category.
They differ only by observation level and do not contradict.

$\mathrm{Desc}(\mathcal{S})$ is not automatically determined as a unique mathematical object.
A concrete theory must specify which descriptions belong to the same system, which structure transformations are considered admissible, and which structures description morphisms must preserve.
Therefore, $\mathrm{Desc}(\mathcal{S})$ is an organizing construction in the framework, rather than an ontological prerequisite for generalized emergence.

\subsection{Category-Theoretic Foundations and Coarse-Graining Functors}

\subsubsection{Categories, Functors, and Structure Preservation}
Category theory expresses structural relations via objects, morphisms, and composition.
It is a natural language for studying transformations between structured relations at different levels; standard references are \cite{MacLane1998, Leinster2014}, and an introduction oriented toward applications outside pure mathematics is \cite{FongSpivak2019}.

Let $\mathcal{C}_\mu$ and $\mathcal{C}_M$ be two description categories.
A coarse-graining functor $Q$ has two parts of action:
\begin{equation}\label{coarse-graining_functor}
Q_0:\mathrm{Ob}(\mathcal{C}_\mu)\to \mathrm{Ob}(\mathcal{C}_M),
\end{equation}
and for each pair of objects, a morphism mapping
\begin{equation}\label{morphism_mapping}
Q_{A,B}:\mathrm{Mor}_{\mathcal{C}_\mu}(A,B)\to \mathrm{Mor}_{\mathcal{C}_M}(Q(A),Q(B)).
\end{equation}
These actions must be compatible with identity morphisms and composition.
Functoriality ensures compatibility with identities and composition; excluding pure renaming requires the separate nontriviality condition below.

This paper does not require all coarse-graining functors to share identical properties.
Some problems require $Q$ to preserve specific structure; others allow $Q$ to forget structure; still others require quotienting or localization.
Admissibility is given by the concrete research domain, not by category theory alone.

\subsubsection{The Class of Admissible Coarse-Grainings}
At the meta-level of $\mathrm{Desc}(\mathcal{S})$, define the class of admissible coarse-grainings:
\begin{equation}\label{admissible_coarse-grainings}
\mathrm{CG}(\mathcal{S})\subseteq \mathrm{Mor}(\mathrm{Desc}(\mathcal{S})).
\end{equation}
For a candidate coarse-graining $Q$, write
\begin{equation}\label{candidate_coarse-graining}
\mathrm{Adm}(Q)\iff Q\in \mathrm{CG}(\mathcal{S}).
\end{equation}
$\mathrm{CG}(\mathcal{S})$ may be specified from different sources; for example, it can require that the coarse-graining preserve some class of designated structure, that it respect the system's compositions and relations, that it be compatible with observation, representation, or experimental operations, that it belong to some allowed variable selection or quotienting scheme, or that it satisfy specific realizability constraints.
These conditions may be expressed in different instances as functorial properties, adjunctions, quotient structures, monotonicity, measurability, or other external constraints.
The key point is that admissibility is a parameter of the theory, not something that must be presupposed as probabilistic, causal, or entropic.

\subsubsection{Nontriviality}
Admissibility alone is not enough to generate emergence.
If $Q$ is merely a renaming, a coordinate change, or a categorical equivalence of the same description, then attribute changes should not be automatically called emergence.
Let $\mathrm{Eq}(\mathcal{S})$ denote the class of coarse-grainings that the theory regards as ``purely descriptive'' (e.g., category isomorphisms or categorical equivalences).
Define
\begin{equation} \label{isomorphisms_categorical_equivalences}
\mathrm{Nontrivial}(Q)\iff Q\in \mathrm{CG}(\mathcal{S})\setminus \mathrm{Eq}(\mathcal{S}).
\end{equation}
This does not claim that every non-equivalent functor must produce emergence;
it claims that emergence at least requires excluding pure descriptive transformations.
More specific nontriviality criteria may require that $Q$ changes distinguishability of objects or morphisms, alters structural degrees of freedom, or reorganizes the hierarchy of description.

Nontriviality must be understood relative to the description norms.
For example, in one theory a forgetting functor may be considered admissible and nontrivial;
in another theory the same forgetting may destroy key structure and thus may not belong to $\mathrm{CG}(\mathcal{S})$.
Therefore, nontriviality is not a single numeric condition independent of theory context.
\subsection{Attribute Assignment and Attribute Lifting}

\subsubsection{Attributes of Descriptions Need Not Be Functors}
Let $\mathcal{V}$ be an attribute space equipped with a comparison relation $\preceq$.
An attribute space can be a partially ordered set, a preordered set, a numerical space, a vector space, a graded set, or a set of propositions.
For description attributes, define an attribute assignment on descriptions and on the morphisms inside them,
\begin{equation}\label{object-level_attribute_assignment}
P:\mathrm{Ob}(\mathrm{Desc}(\mathcal{S}))\to (\mathcal{V},\preceq).
\end{equation}
Thus $P(\mathcal{C}_\mu)$ and $P(\mathcal{C}_M)$ are the attribute values of the microscopic and macroscopic descriptions, respectively.
A description is a category, so its value may be computed from any part of its data: a property of its objects, of its morphisms, or of a designated mechanism that it carries.
Which morphism is designated is part of the description, so $P$ remains a function of the description; the macroscopic value is then computed from $Q(f)$.
The instances of Section~\ref{sec:dynamical-instances} evaluate a designated mechanism, because effective information and dynamical reversibility are properties of transition morphisms rather than of state objects.

One should carefully distinguish attribute assignments from attribute functors.
In general, $P$ only assigns attribute values and does not specify how those values vary along morphisms between descriptions.
Thus, one cannot unconditionally write $P$ as a functor from $\mathrm{Desc}(\mathcal{S})$ to some attribute category.
If a particular attribute indeed induces morphism-level action and preserves identities and composition, then $P$ may be strengthened to an attribute functor.
This is useful additional structure but not a foundation requirement for generalized emergence.

\subsubsection{Attribute Order and Lifting}\label{Order_and_Lifting}
A macroscopic ``lift'' of an attribute does not necessarily mean the numeric value increases.
Let $\prec$ denote the strict improvement relation on the attribute space.
Attribute lifting is defined by
\begin{equation}\label{Attribute_lifting}
\mathrm{Lift}_P(Q)\iff P(\mathcal{C}_\mu)\prec P(\mathcal{C}_M),\qquad Q:\mathcal{C}_\mu\to\mathcal{C}_M.
\end{equation}
For a designated mechanism $f_\mu\in\mathrm{Mor}(\mathcal{C}_\mu)$, the same relation compares $P(f_\mu)$ with $P\bigl(Q(f_\mu)\bigr)$, so the macroscopic value is taken at the functorial image and describes the same mechanism only when $Q(f_\mu)$ is the macroscopic mechanism.
Here $\mathcal{C}_M$ is the target description of $Q$, not the image of $\mathcal{C}_\mu$ as an object inside its own category.
With the microscopic description fixed, changing $Q$ may change the target description and therefore the attribute comparison.
If only non-decrease is required, use the non-strict relation $\preceq$.
When attributes represent errors, redundancy, or leakage, researchers may reverse the convention so that ``smaller is better'' while still using the same order relation to express lifting.

Numeric attributes are a special case.
If $P:\mathrm{Ob}(\mathrm{Desc}(\mathcal{S}))\to \mathbb{R}$,
one may define attribute gain
\begin{equation}\label{attribute_gain}
E_P(Q)=P(\mathcal{C}_M)-P(\mathcal{C}_\mu).
\end{equation}
Under a convention ``larger is better,'' $E_P(Q)>0$ indicates strict lifting.
However, the generalized theory does not require attributes to be numeric.

\subsubsection{Multiple Attributes and Partial Orders}
A system often possesses multiple attributes simultaneously.
Let 
\begin{equation}\label{multi-attribute}
P=(P_1,P_2,\dots,P_n)
\end{equation}
be a multi-attribute evaluation.
Then the attribute space naturally has only a partial order, not necessarily a total order.
In this case, a macroscopic description may improve some dimensions, keep others unchanged, or form Pareto improvements.
This avoids collapsing ``emergence'' to the maximization of a single indicator.
In applications one may choose a main attribute, a weighted order, a partial order, or a set of mutually irreducible attributes.
Different choices may yield different emergence judgments.
This is not a deficiency of the framework; it reflects that emergence is explicitly defined relative to research purposes.

\subsection{The Generalized Emergence Criterion}\label{sec:emergence-criterion}

\subsubsection{Definition of Generalized Emergence}
\begin{definition}[Generalized emergence]\label{def:generalized-emergence}
Fix a system $\mathcal{S}$, a microscopic description $\mathcal{C}_\mu$, an admissible coarse-graining class, and an attribute $P$ with its improvement order.
A functor $Q:\mathcal{C}_\mu\to\mathcal{C}_M$ witnesses generalized emergence relative to $P$ precisely when
\begin{equation}\label{generalized_emergence}
\mathrm{Emergence}_P(Q)\iff
\mathrm{Adm}(Q)\wedge \mathrm{Nontrivial}(Q)\wedge\mathrm{Lift}_P(Q).
\end{equation}
\end{definition}
When $P$ is the morphism component read off a designated mechanism, the same criterion applies with the mechanism-level comparison of Section~\ref{Order_and_Lifting} in place of $\mathrm{Lift}_P(Q)$.
Admissibility excludes transformations outside the allowed class, nontriviality excludes pure renaming, and attribute lifting supplies the improvement being tested.
Thus the definition concerns an attribute of a description or of a designated morphism inside it, not an attribute of the relation between descriptions.

\subsubsection{Dependence on the Coarse-Graining Functor}\label{sec:functor-dependence}
Generalized emergence is functor-dependent, not unconditional: with $\mathcal{C}_\mu$ and $P$ fixed, different $Q$ may or may not lift $P$, so an emergence claim must exhibit its witness $Q$. “Admitting emergence” means such a $Q$ exists, not that all coarse-grainings work. This is observer-dependence in the representational sense — different $Q$ retain/merge distinctions differently — but once scheme, attribute, and admissibility are fixed, the judgment is determined, not subjective.

\subsubsection{Basic Propositions}
\begin{proposition}[Pure descriptive coarse-grainings do not produce generalized emergence]
If $Q\in \mathrm{Eq}(\mathcal{S})$ and attribute $P$ is invariant on $\mathrm{Eq}(\mathcal{S})$, then $Q$ does not satisfy $\mathrm{Emergence}_P(Q)$.
\end{proposition}
\begin{proof}
Since $Q\in \mathrm{Eq}(\mathcal{S})$, that is, $Q$ is purely descriptive (a category isomorphism or a categorical equivalence of the same description), $\mathrm{Nontrivial}(Q)$ fails.
Therefore the conjunctive emergence criterion does not hold.
\end{proof}

\begin{proposition}[Composition of attribute lifts]
Assume two consecutive admissible coarse-grainings
\begin{equation}\label{consecutive_admissible_coarse-grainings}
    \mathcal{C}_0 \xrightarrow{Q_1} \mathcal{C}_1 \xrightarrow{Q_2} \mathcal{C}_2,
\end{equation}
and assume the attribute order $\prec$ is transitive.
If $P(\mathcal{C}_0)\prec P(\mathcal{C}_1)$ and $P(\mathcal{C}_1)\prec P(\mathcal{C}_2)$, and if the composed coarse-graining $Q_2\circ Q_1$ is admissible and nontrivial, then
\begin{equation}\label{composed_coarse-graining}
P(\mathcal{C}_0)\prec P(\mathcal{C}_2),\qquad \mathrm{Emergence}_P(Q_2\circ Q_1)
\end{equation}
holds.
\end{proposition}
\begin{proof}
By transitivity of $\prec$, we get $P(\mathcal{C}_0)\prec P(\mathcal{C}_2)$.
Together with admissibility and nontriviality of the composed coarse-graining, these inequalities satisfy the three conditions of Definition~\ref{def:generalized-emergence}.
\end{proof}

\noindent
These conditions indicate that attribute lifting can compose along the hierarchy,
but one cannot conclude that any two coarse-grainings composed together must necessarily produce emergence.
Whether the composite is still admissible and still nontrivial must be verified separately.

\subsection{Selecting an Optimal Coarse-Graining}\label{sec:optimal-coarse-graining}
An emergence judgment tests a given $Q$; it does not by itself select one.
To select a coarse-graining, fix the microscopic description $\mathcal{C}_\mu$, the allowed macroscopic descriptions, and the admissibility constraints before optimization.
Let $\mathcal{Q}(\mathcal{C}_\mu)$ be the resulting set of admissible, nontrivial functors with source $\mathcal{C}_\mu$.
For each candidate $Q$, write $\mathcal{C}_M^Q$ for its target description. Constraints may fix the macro resolution or require preservation of designated structure. These constraints prevent an unacceptable collapse from becoming optimal merely because it raises the chosen attribute.

For a real-valued description attribute $P$ with ``larger is better,'' define the set of optimal coarse-grainings by
\begin{equation}\label{optimal_coarse_graining}
\mathcal{Q}_P^\star(\mathcal{C}_\mu)
=\underset{Q\in\mathcal{Q}(\mathcal{C}_\mu)}{\operatorname{arg\,max}}\,
P(\mathcal{C}_M^Q).
\end{equation}
Since $P(\mathcal{C}_\mu)$ is fixed, this also maximizes the attribute gain $E_P(Q)$.
If smaller values are better, use minimization instead.
For vector attributes with a partial order, one generally obtains a set of Pareto-optimal candidates; a scalar objective or a specified preference rule is needed to select among them.

Under the observer-dependent interpretation of Section~\ref{sec:functor-dependence}, maximizing the macroscopic attribute $P$ is equivalent to seeking an optimal observational perspective among those represented by the candidate functors.
This optimality is relative to the selected attribute, its improvement order, and the constraints defining $\mathcal{Q}(\mathcal{C}_\mu)$.
Changing the attribute or the constraints may select a different perspective; there is no attribute-independent or constraint-independent best view of the system.
Selecting an optimal coarse-graining can therefore reduce, to some extent, the observer dependence of emergence judgments by replacing an arbitrary choice of perspective with a shared optimization criterion, while retaining dependence on the chosen attribute and constraints.

Existence and uniqueness are separate requirements.
A nonempty finite candidate set guarantees attainment of a maximum; in an infinite class a supremum need not be attained.
Even when a maximum exists, several functors may attain it.
Only when $\mathcal{Q}_P^\star(\mathcal{C}_\mu)=\{Q^\star\}$
%\begin{equation}\label{unique_optimal_coarse_graining}
%\mathcal{Q}_P^\star(\mathcal{C}_\mu)=\{Q^\star\}
%\end{equation}
does maximization uniquely determine the coarse-graining in that candidate set.
A unique optimal target description is not sufficient: distinct functors can have the same target and therefore the same value of $P$.
%If macro relabelings are regarded as the same scheme, optimization should be performed over the corresponding equivalence classes, assuming the attribute and constraints are invariant under those relabelings; uniqueness then means a single optimal class. An explicit tie-breaking rule can select one representative, but this is a modeling convention, not uniqueness implied by the attribute alone.

Finally, an optimal candidate witnesses generalized emergence only if its macroscopic attribute strictly improves on $P(\mathcal{C}_\mu)$.
Optimality is comparison among allowed coarse-grainings; emergence is comparison with the microscopic description.
The former does not imply the latter.
For causal emergence, maximizing macroscopic effective information exemplifies this selection principle \cite{Hoel2013}; its constraints and limitations are discussed in Section~\ref{sec:dynamical-instances}.

\subsection{Generalized Emergence in Dynamical Systems}\label{sec:dynamical-framework}
The base definition of generalized emergence does not require a dynamical semantics.
However, once a system description chooses to represent transformations, evolution, or stochastic processes by morphisms, we obtain an important special case: generalized emergence in dynamical systems.

Here we must distinguish two operations that, in general, are not the same:
state coarse-graining maps microscopic states/distributions to macroscopic states/distributions;
dynamical coarse-graining maps microscopic evolution rules to macroscopic evolution rules.
Their compatibility, not the requirement that both must be provided by a single functor, is the basic requirement for macroscopic dynamical descriptions.
Computational mechanics, causal emergence and approximate dynamical reversibility are discussed using this dynamical representation in Section~\ref{sec:dynamical-instances}.
%The distinction between $\Phi$ and $\Psi$ is a modeling distinction, not a replacement for the functor required by Definition~\ref{def:generalized-emergence}: a pair of maps alone does not establish that a coarse-graining functor exists.
%A strict categorical emergence claim must identify the relevant description categories and functor; approximate effective models remain application-level constructions.

\subsubsection{Dynamical Endogenization Inside a Category}
The categorical emergence skeleton does not require the system to have a dynamical semantics.
But once the chosen semantic category represents transitions/evolution/stochastic processes by morphisms, dynamics can be endogenized as morphisms in that category rather than additional objects outside the category.
%Objects represent state spaces;
%concrete states are generalized elements from the unit object to the state space;
%dynamics are morphisms between state spaces.
%Objects themselves do not participate in composition; real evolution is composition of a generalized element (state) with a dynamical morphism.

For Markov dynamics, an appropriate semantic category can be a Markov category (or an equivalent Kleisli-type probability category) \cite{Fritz2020}.
In a typical Markov category $\mathbf{Mark}$:
\begin{itemize}
  \item Objects are state spaces, e.g., measurable spaces or finite sets $X$;
  \item Morphisms are Markov kernels $K: X \rightsquigarrow Y$;
  \item A discrete-time Markov dynamics is an endomorphism $K: X \rightsquigarrow X$;
  \item A probability distribution $p:I\rightsquigarrow X$ is a morphism from the unit object $I$ to $X$, i.e., a generalized element;
  \item Composition $K\circ p: I\rightsquigarrow X$ gives the pushforward of the distribution after applying dynamics.
\end{itemize}

For finite state spaces $X$, Markov kernels also induce affine maps on probability simplices, $K^\ast:\Delta(X)\to \Delta(X)$.
If the research goal requires treating distribution spaces as objects, one may move to a convex category.
This is an alternative semantic implementation, not a necessary choice of the basic framework.

\subsubsection{State Coarse-Graining and Dynamical Coarse-Graining}
Let microscopic and macroscopic state spaces be $X_\mu$ and $X_M$, with corresponding unit objects $I_\mu$ and $I_M$.
State coarse-graining is a mapping defined on generalized elements:
\begin{equation}\label{State_coarse-graining}
    \Phi_X:\mathrm{Hom}_{\mathcal{C}_\mu}(I_\mu,X_\mu)\to \mathrm{Hom}_{\mathcal{C}_M}(I_M,X_M),\qquad x\mapsto \Phi_X(x).
\end{equation}
It specifies how each microscopic state or microscopic distribution is represented as a macroscopic state or macroscopic distribution.

For any related microscopic transformation $f:X_\mu\to Y_\mu$, dynamical coarse-graining is an independent mapping
\begin{equation}\label{dynamical_coarse-graining}
\Psi_{X,Y}:\mathrm{Hom}_{\mathcal{C}_\mu}(X_\mu,Y_\mu)\to \mathrm{Hom}_{\mathcal{C}_M}(X_M,Y_M),\qquad
f\mapsto \Psi_{X,Y}(f).
\end{equation}
The maps $\Phi$ and $\Psi$ differ in domain, semantics, and modeling origin:
$\Phi$ typically comes from observation, encoding, aggregation, or distribution pushforward;
$\Psi$ typically comes from model reduction, parameter fitting, dynamical projection, or constructions from other evolution rules.
Therefore, in general settings one should not expect that $\Phi$ and $\Psi$ are restrictions of a single mapping on one morphism type.

\subsubsection{Dynamical Closure and a Strict Functorial Special Case}
The basic closure condition relating state and dynamical coarse-graining is:
for every relevant state $x:I_\mu\to X_\mu$ and every dynamical morphism $f:X_\mu\to Y_\mu$, we require
\begin{equation}\label{dynamical_morphismrequire}
\Phi_Y(f\circ x)=\Psi_{X,Y}(f)\circ \Phi_X(x).
\end{equation}
%6.1
This expresses that ``evolve in the micro level then coarse-grain'' equals ``coarse-grain the state then apply the macroscopic dynamics''.
It is a compatibility condition between $\Phi$ and $\Psi$, not guaranteed by $\Phi$ or $\Psi$ alone.

For an endomorphism $K_\mu:X_\mu\rightsquigarrow X_\mu$, define the macroscopic dynamics
\begin{equation}
K_M:=\Psi_{X,X}(K_\mu).
\end{equation}
Then Eq.~\ref{dynamical_morphismrequire} becomes
\begin{equation}\label{macroscopic_morphismrequire}
\Phi_X(K_\mu\circ x)=K_M\circ \Phi_X(x).
\end{equation}
%6.2

In a finite Markov chain, if state coarse-graining is induced by a deterministic mapping $q:X_\mu\to X_M$, then distribution coarse-graining is the pushforward $q^\ast$.
The existence condition for $K_M$ becomes $K_M\circ q^\ast=q^\ast\circ K_\mu$,
which is a standard form of lumpability (dynamical factorization) \cite{KemenySnell1960}.
Note that $q^\ast\circ K_\mu$ still has a domain in the micro-distribution space, so it cannot directly be treated as the macroscopic dynamics.
Only when the factorization condition holds does one obtain a genuinely defined macroscopic $K_M$ on the macroscopic space.

A unified functorial description is a further and strong strict special case:
if there exists a unit-preserving (or structure-isomorphic) functor $Q:\mathcal{C}_\mu\to \mathcal{C}_M$ such that
\begin{equation}\label{structure-isomorphic}
X_M=Q_0(X_\mu),\quad Y_M=Q_0(Y_\mu),
\end{equation}
and
\begin{equation}\label{structure-isomorphic_phi}
\Phi_X(x)=Q_1(x),\qquad \Psi_{X,Y}(f)=Q_1(f),
%\tag{6.3}
\end{equation}
then equation \ref{dynamical_morphismrequire} follows automatically from preservation of composition by the functor:
\begin{equation}\label{composition_functor}
Q_1(f\circ x)=Q_1(f)\circ Q_1(x).
\end{equation}
This strict special case should not be viewed as the general definition of dynamical coarse-graining; it only describes a special situation where states and dynamics can be uniformly handled by one structure-preserving mapping.

If strict closure holds for one-step dynamics and $\Psi$ preserves composition along evolution paths formed by the relevant dynamical morphisms, identity morphisms, and finite compositions,
then this compatibility propagates to all finite-step evolutions.
This indicates that dynamical closure covers that evolution range.

\subsubsection{Approximate Closure}
In applications, state coarse-graining $\Phi$ and macroscopic effective dynamics $K^{\mathrm{eff}}_M$ often come from finite observations, parameterized models, or data estimation,
so they need not strictly satisfy Eq.~\ref{macroscopic_morphismrequire}.
If a distance/divergence/cost function $d$ is given on the macroscopic state space, define the closure defect by
\begin{equation}\label{closure_defect}
\varepsilon(\Phi,K^{\mathrm{eff}}_M;K_\mu)=
\sup_{x\in A}\,
d\Bigl(\Phi_X(K_\mu\circ x),\,K^{\mathrm{eff}}_M\circ \Phi_X(x)\Bigr),
%\tag{6.4}
\end{equation}
where $A$ is the considered class of microscopic states or distributions.
If $\varepsilon\le \delta$ for a pre-given tolerance $\delta$, then the state--dynamics coarse-graining is called approximately closed at tolerance $\delta$.
If $K^{\mathrm{eff}}_M$ ranges over an allowed macroscopic model class $D_M$, one may further take the infimum/supremum over that class to evaluate the best achievable closure within the model class.

Therefore, approximate closure is not a modification of functor axioms and does not need a strictly induced macroscopic dynamics as a reference.
It first characterizes the mismatch between $\Phi$ and the effective dynamical coarse-graining.
A strict functorial special case provides a zero-defect benchmark model.
In this framework, closure defects serve as admissibility or model-accuracy constraints on a dynamical coarse-graining, not as attributes that independently witness emergence.

\section{Examples}
\subsection{Existing Special Cases in Dynamical Systems}\label{sec:dynamical-instances}
This section rewrites existing theories uniformly as special cases of the dynamical representation in Section~\ref{sec:dynamical-framework}.
Each instance should specify microscopic and macroscopic state spaces, state coarse-graining $\Phi$, microscopic dynamical morphisms $K_\mu$, and dynamical coarse-graining or effective macroscopic dynamics $\Psi(K_\mu)$ (or $K^{\mathrm{eff}}_M$), together with strict or approximate closure relations between them.
Only when there exists an extra unit-preserving functor can $\Phi$ and $\Psi$ be written simultaneously as morphism actions of that functor.
The three theories differ in their coarse-graining construction and evaluation attributes.
Their numerical comparisons evaluate descriptions with designated mechanisms.
To obtain strict instances of Definition~\ref{def:generalized-emergence}, one must also supply an admissible, nontrivial functor $Q$ between those descriptions; numerical improvement alone is insufficient.
Because the attributes used here are read off the mechanism, a strict instance additionally requires $Q$ to map the microscopic mechanism to the macroscopic one.
Table~\ref{tab:instances} summarizes this correspondence.

\paragraph{Computational mechanics.}
Crutchfield and colleagues reconstruct the predictive structure of a stationary stochastic process by grouping histories that induce the same conditional distribution of the entire future; each equivalence class is a \emph{causal state} \cite{CrutchfieldYoung1989,Crutchfield1994}.
Thus, the coarse-graining merges predictively equivalent histories into causal states, rather than arbitrarily merging distinct causal states.
The state map $\Phi$ is this quotient, and its compatibility with successive symbol updates induces the causal-state dynamics, represented by the $\epsilon$-machine.
A coarse-graining functor $Q$ between the corresponding dynamical descriptions must implement this quotient together with the induced updates.
For a description $\mathcal{C}$ with a designated predictive state $R_{\mathcal{C}}$, we may choose predictive efficiency as the attribute\cite{Crutchfield1994}
\begin{equation}\label{predictive_efficiency}
P_{\mathrm{pred}}(\mathcal{C})=
\frac{I(R_{\mathcal{C}};X^+)}{H(R_{\mathcal{C}})},
\qquad 0<H(R_{\mathcal{C}})<\infty,
\end{equation}
where $X^+$ denotes the future, $H$ is Shannon entropy, and $I$ is mutual information.
For the causal-state representation $R_{\mathcal{C}_M}=S$, predictive sufficiency gives $I(S;X^+)=\mathbf{E}=I(X^-;X^+)$, the excess entropy, where $X^-$ denotes the past; $H(S)=C_{\mathrm{stat}}$ is the statistical complexity.
Its predictive efficiency is therefore $\mathbf{E}/C_{\mathrm{stat}}$: predictive information per bit of stored memory.
When a redundant predictive representation is quotiented into causal states, the predictive information is retained while the memory entropy cannot increase; efficiency rises strictly only if that entropy decreases and the retained predictive information is positive.
Such a rise witnesses generalized emergence when $Q$ is admissible and nontrivial; causal-state reconstruction alone does not guarantee a strict lift.

\subsubsection{Causal Emergence: A Causal Dynamical Special Case}
\label{sec:causal_emergence}
Hoel, Albantakis, and Tononi's theory of causal emergence uses effective information to quantify cause-effect power across levels and compares the micro and macro layers via coarse-graining \cite{Hoel2013}.
When the macro level's effective information exceeds the micro level's, the system is judged to have causal emergence \cite{Hoel2013,Yuan2024}.

In the framework of Section~\ref{sec:dynamical-framework}, microscopic and macroscopic state spaces are objects; concrete states and probability distributions are generalized elements from the unit object; the microscopic mechanism is the relevant dynamical morphism $K_\mu$. The state coarse-graining $\Phi$ acts on states and on probability distributions: any distribution over microscopic states is pushed forward to a distribution over macroscopic states.

What the theory actually needs is the coarse-graining of the transition matrix that this state coarse-graining induces.
The macroscopic mechanism $K_M$ is obtained by aggregating the entries of $K_\mu$ over the macroscopic blocks --- for instance by block averaging --- or by an independent dynamical coarse-graining $\Psi(K_\mu)$ fitted to an empirical model; the two agree when the closure relation of Section~\ref{sec:dynamical-framework} holds.
Interventions enter only in the definition of the attribute: effective information is computed from interventional distributions in the original theory \cite{Hoel2013}, whereas the coarse-graining itself is applied to ordinary state distributions. When the closure relation holds, the macro mechanism faithfully represents micro evolution on the considered state classes; functorial strictification is a sufficient but not necessary way to realize this relationship.

Effective information provides $P_{\mathrm{EI}}$, which is a property of the mechanism rather than of the state object\cite{Zhang2025}.
Its increase under the state--dynamics coarse-graining is an application-level attribute comparison; it witnesses generalized emergence in the present sense only after an admissible, nontrivial functor $Q$ between the two descriptions is specified, and then the comparison is between $K_\mu$ and its functorial image, so $Q(K_\mu)=K_M$ is required.

The selection of the coarse-graining is also part of this theory.
In the original theory, the macroscopic partition is chosen by maximizing the effective information of the macroscopic dynamics \cite{Liu2024}, and the defect that this choice may introduce into the closure relation is not part of the objective; this has been criticized as letting a distortion in the map obscure the territory \cite{EberhardtLee2022}.
Later work corrects the objective by requiring, in addition to a maximal macroscopic effective information, that macroscopic states predict microscopic states up to a prescribed accuracy \cite{Zhang2022,Yang2025}.
That requirement is a variant of the defect condition \eqref{closure_defect}: the micro-to-macro defect asks whether the coarse-grained dynamics can be reproduced at the macroscopic level, the macro-to-micro accuracy asks whether the coarse-graining can be inverted well enough for a macroscopic state to stand in for the microscopic ones it summarizes, and a usable coarse-graining needs both defects to be small.

\subsubsection{Dynamical Reversibility Emergence: A Markov Dynamical Special Case}
This instance keeps the skeleton of Section~\ref{sec:causal_emergence} and replaces only the measure: effective information is replaced by state-averaged approximate dynamical reversibility. A finite Markov chain is rarely invertible, so instead of asking whether the transition map can be undone, Zhang et al.\ read the dynamics off the singular-value spectrum of the transition probability matrix $T_\mu=U\Sigma V^{\top}$, with ordered singular values $\sigma_1\ge\cdots\ge\sigma_N\ge 0$ \cite{Zhang2025,liu2025singular}. As defined, the theory needs no coarse-graining at all: the $\alpha$-ordered approximate dynamical reversibility is $\Gamma_\alpha=\sum_{i=1}^{N}\sigma_i^{\alpha}$, its state-averaged form over the $r$ largest singular directions is 
\begin{equation}
P_{\mathrm{rev}}\equiv\gamma_\alpha(r)=\frac{1}{r}\sum_{i=1}^{r}\sigma_i^{\alpha},
\end{equation}
and for a threshold $\epsilon\ge 0$ the degree of emergence is the improvement obtained by discarding the singular values below $\epsilon$ and re-averaging over the $r_\epsilon=\max\{i:\sigma_i>\epsilon\}$ retained directions,
\begin{equation}\label{eq:rev-emergence}
\Delta\Gamma_\alpha(\epsilon)=\gamma_\alpha(r_\epsilon)-\gamma_\alpha(N),
\end{equation}
with emergence declared when this quantity is positive. Both values are computed from the single microscopic matrix: no macroscopic state space, no macroscopic mechanism, and no coarse-graining functor appear; the authors state that the quantification is independent of the choice of coarse-graining method \cite{Zhang2025}.

The spectral truncation can nevertheless be read as an approximate coarse-graining: keeping the $r_\epsilon$ leading singular directions compresses the mechanism to
\begin{equation}\label{svd_coarse_graining}
T_\mu\longmapsto\widetilde{T}=U_{r_\epsilon}\Sigma_{r_\epsilon}V_{r_\epsilon}^{\top},
\end{equation}
so that coarse-graining acts on the mechanism rather than on the states, the ``macroscopic'' object being the subspace spanned by the leading singular vectors rather than a set of macro-states. The reading is approximate precisely because $\widetilde{T}$ is a linear object: its entries need not be nonnegative and its rows need not sum to one, so it is not a transition probability matrix, not a Markov kernel in the sense of Section~\ref{sec:dynamical-framework}, and the truncation, defined one mechanism at a time from that mechanism's own spectrum, does not extend to a coarse-graining functor.

The framework locates the three roles separately. The attribute $P_{\mathrm{rev}}$ is computed from a designated mechanism, exactly the situation allowed in Section~\ref{Order_and_Lifting}: a property of the transition morphism, not of the state object. The quantity $\Delta\Gamma_\alpha(\epsilon)$ belongs to the logic of Section~\ref{sec:optimal-coarse-graining}: it measures the improvement achievable by an optimal coarse-graining that projects probability mass onto the leading singular directions without that strategy being explicitly constructed \cite{Zhang2025}, so the spectrum certifies that an admissible, nontrivial $Q$ with positive attribute gain exists rather than exhibiting the witness. The explicit witness is supplied by the constructive algorithm of \cite{Zhang2025}: the rows of $T_\mu$ are projected onto the leading right singular vectors, clustered into a genuine partition of the microscopic states, and the stationary flow is aggregated over this partition into a macroscopic TPM. From there the dynamical skeleton of Section~\ref{sec:dynamical-framework} applies verbatim, with $Q(T_\mu)=T_M$; the spectrum merely guides the choice of $Q$ instead of an exhaustive partition search, and since the construction approximately maximizes macroscopic effective information \cite{Zhang2025}, the instance remains comparable with the causal-emergence instance of Section~\ref{sec:causal_emergence}.

\begin{table}[htbp]
\centering
\caption{The three dynamical theories in the state--dynamics representation of Section~\ref{sec:dynamical-framework}. Each row gives the state coarse-graining, the construction of the macroscopic dynamics, and the attribute being compared.}
\label{tab:instances}
\footnotesize
\begin{tabular}{>{\raggedright\arraybackslash}p{0.2\textwidth}|>{\raggedright\arraybackslash}p{0.4\textwidth}|>{\raggedright\arraybackslash}p{0.25\textwidth}}
\hline
\textbf{Theory instance} &
\textbf{State--dynamics representation in Section~\ref{sec:dynamical-framework}} &
\textbf{Lifted properties} \\
\hline
Computational mechanics
& $\Phi$ groups predictively equivalent histories into causal states; the compatible symbol updates induce the $\epsilon$-machine dynamics.
& predictive efficiency $P_{\mathrm{pred}}$, equal to excess entropy divided by statistical complexity for the causal-state representation \\
\hline
Causal emergence
& $\Phi$ coarse-grains states and state distributions; the induced coarse-graining of the transition matrix gives the macroscopic mechanism, from which effective information is computed.
& effective information/cause-effect power ($P_{\mathrm{EI}}$) \\
\hline
Dynamical reversibility (SVD)
& $\Phi$ pushes forward state distributions; leading singular directions of the transition matrix guide the construction of the macroscopic mechanism.
& average approximate dynamical reversibility ($P_{\mathrm{rev}}$) \\
\hline
\end{tabular}
\end{table}

The table's meaning is structural correspondence, not indicator identity.
Predictive efficiency, effective information, and average approximate dynamical reversibility address different questions; the correspondence does not identify their values.

\subsection{Non-Dynamical Illustrative Example}
% 建议替换 3.2 节开头的概述段落，以匹配更新后的两个案例
In addition to generalized emergence based on dynamics, we also present two non-dynamical examples below. The first compares two coarse-graining outcomes for
closed polygons: averaging groups of vertices reduces both symmetry
distances in one case, but produces mixed changes in the other. The second
compares two network structures under community-based coarse-graining:
cohesion indicators increase for a weakly coupled modular network, whereas
they decrease for a network whose communities form a ring. Together, these
examples illustrate that attribute lifting is conditional on both the
system's structure and the chosen coarse-graining.

\subsubsection{Symmetry Emergence of Polygons}\label{sec:polygon-example}
\begin{figure}[!htp]
    \centering
    \includegraphics[width=0.95\textwidth,trim=0 70 330 20, clip]{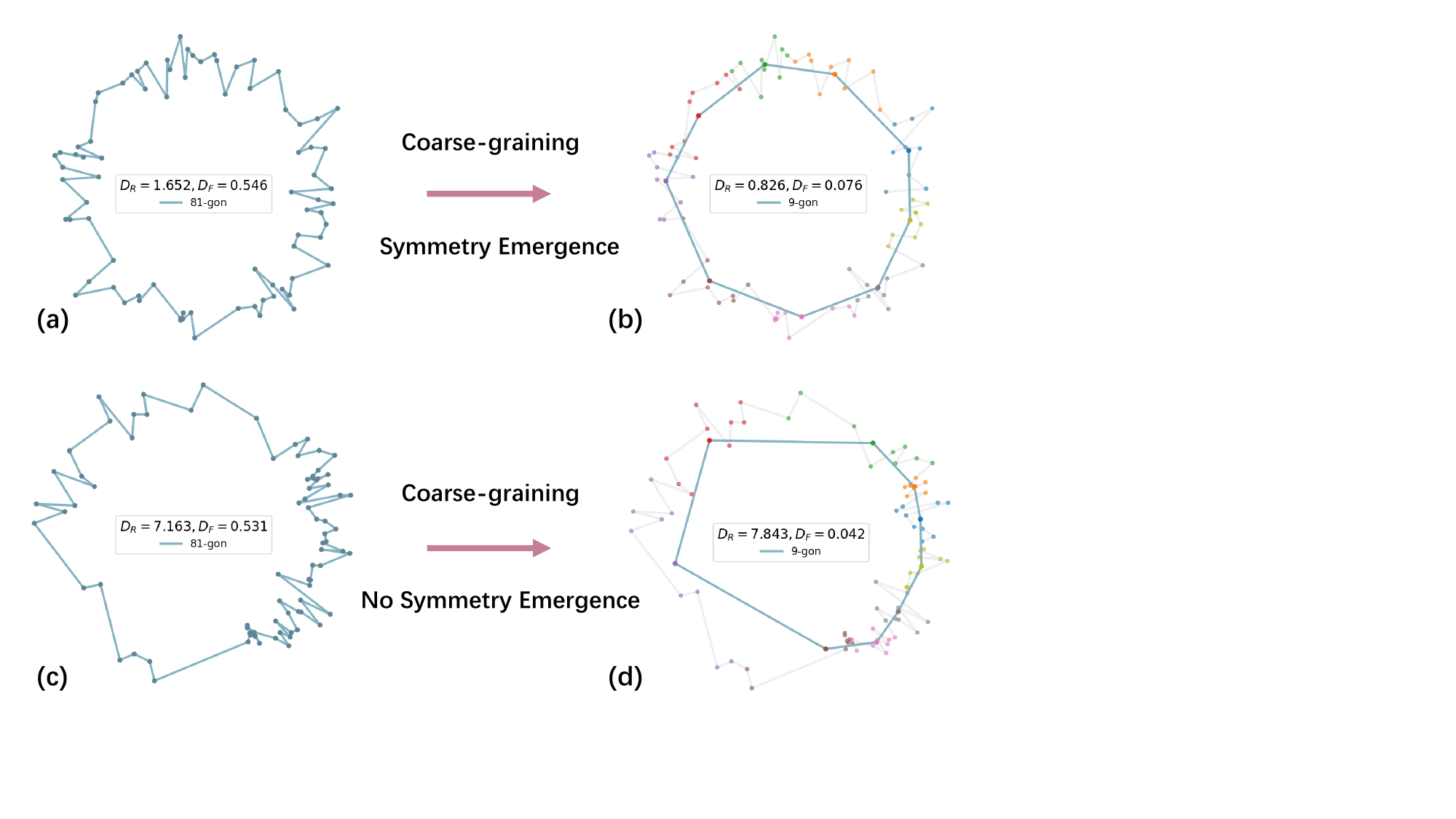}
    \caption{\textbf{Symmetry distances can either decrease or change in
    opposite directions under coarse-graining.} We apply coarse-graining strategies that group nine adjacent same-colored nodes and calculate their average position, thereby transforming the microscopic 81-gons into macroscopic 9-gons.
    The top row, panels (a)--(b), shows a comparatively uniform imperfect
    81-gon and the 9-gon obtained by averaging each group of nine consecutive
    vertices. Both symmetry distances decrease, and the 81-gon system exhibits symmetry emergence under coarse-graining. The bottom row, panels
    (c)--(d), shows a less uniform imperfect 81-gon and its coarse-grained
    9-gon; here the rotational and dihedral symmetry distances change in
    opposite directions, so the 81-gon system exhibits no symmetry emergence.}
    \label{fig:schematic_diagram}
\end{figure}

In this section, planar polygons, together with their vertex sequences and incidence structures,
are descriptions of the systems under study. We apply $\mathcal{C}_N$ for the
polygons with $N$ vertices and consider two description levels,
$\mathcal{C}_{81}$ and $\mathcal{C}_{9}$. Coarse-graining averages each group
of nine consecutive vertices into one vertex, giving the functor
\begin{equation}
\mathcal{C}_{81}\xrightarrow{Q_{81\to9}}\mathcal{C}_{9}.
\end{equation}
Thus, this example compares the
microscopic 81-gon with its macroscopic 9-gon.

We characterize polygonal symmetry by continuous symmetry measures \cite{zabrodsky1992continuous} using the distances of each node to the nearest $C_N$-rotationally symmetric and $\sigma$-dihedral symmetric configurations, denoted by
$D_R(\mathcal{C})$ and $D_F(\mathcal{C})$, respectively. Both distances
vanish for configurations with the corresponding symmetry, so smaller values
indicate greater symmetry. We therefore take
\begin{equation}
P=(-D_R,-D_F),
\end{equation}
and read $P(\mathcal{C})\prec P(\mathcal{C}')$ as ``both distances are strictly smaller'', that is, as strict improvement in symmetry.
This is the convention of Section~\ref{Order_and_Lifting} for attributes such as errors and distortions, applied without changing the definition of lifting. In particular,
\begin{equation}
\mathrm{Lift}_{P}(Q_{81\to9})
\iff
D_R(\mathcal{C}_{9})<D_R(\mathcal{C}_{81})
\ \text{and}\
D_F(\mathcal{C}_{9})<D_F(\mathcal{C}_{81}).
\end{equation}

For the comparatively uniform polygon in Figure~\ref{fig:schematic_diagram}
(a)--(b), the measured distances are listed in the two columns under Case~1
in Table~\ref{tab:Symmetrydistances}.
%\begin{equation}
%\begin{array}{lcc}
% & \mathcal{C}_{81} & \mathcal{C}_{9}\\[2pt]
%D_R & 1.652 & 0.826\\
%D_F & 0.546 & 0.076
%\end{array}
%\end{equation}
Both distances decrease, so
$P(\mathcal{C}_{81})\prec P(\mathcal{C}_{9})$ and the coarse-graining
satisfies $\mathrm{Lift}_{P}(Q_{81\to9})$. Relative to the selected
attributes, this case exhibits symmetry emergence: both rotational and
dihedral symmetry are more strongly expressed at the macroscopic level.

\begin{table}[htbp]
\centering
\begin{tabular}{l|cc|cc}
\hline
  & \multicolumn{2}{c|}{Case 1 (symmetry emergence)}
  & \multicolumn{2}{c}{Case 2 (no symmetry emergence)} \\\cline{2-3}\cline{4-5}
  & $\mathcal{C}_{81}$ & $\mathcal{C}_{9}$
  & $\mathcal{C}_{81}$ & $\mathcal{C}_{9}$ \\
\hline
$D_R$ & 1.652 & 0.826 & 7.163 & 7.843 \\
$D_F$ & 0.546 & 0.076 & 0.531 & 0.042 \\
\hline
\end{tabular}
\caption{Rotational and dihedral symmetry distances of polygons in two cases.}
\label{tab:Symmetrydistances}
\end{table}

The second polygon, shown in Figure~\ref{fig:schematic_diagram} (c)--(d),
has a less uniform distribution of vertices. Its measured distances are listed in the two columns under Case~2
in Table~\ref{tab:Symmetrydistances}.
%\begin{equation}
%\begin{array}{lcc}
% & \mathcal{C}_{81} & \mathcal{C}_{9}\\[2pt]
%D_R & 7.163 & 7.843\\
%D_F & 0.531 & 0.042
%\end{array}
%\end{equation}
Here the rotational symmetry distance increases, while the dihedral symmetry
distance decreases. Consequently, neither
$P(\mathcal{C}_{81})\prec P(\mathcal{C}_{9})$ nor the reverse componentwise
improvement holds, and the coarse-graining does not satisfy
$\mathrm{Lift}_{P}(Q_{81\to9})$. This comparison shows that symmetry emergence
is not guaranteed by coarse-graining alone: its outcome depends on the
geometry of the microscopic configuration and on the attributes used to
evaluate symmetry.

\subsubsection{Cohesion Emergence of Complex Networks}\label{sec:community-example}
\begin{figure}[!htp]
    \centering
    \includegraphics[width=0.95\textwidth,trim=10 50 180 10, clip]{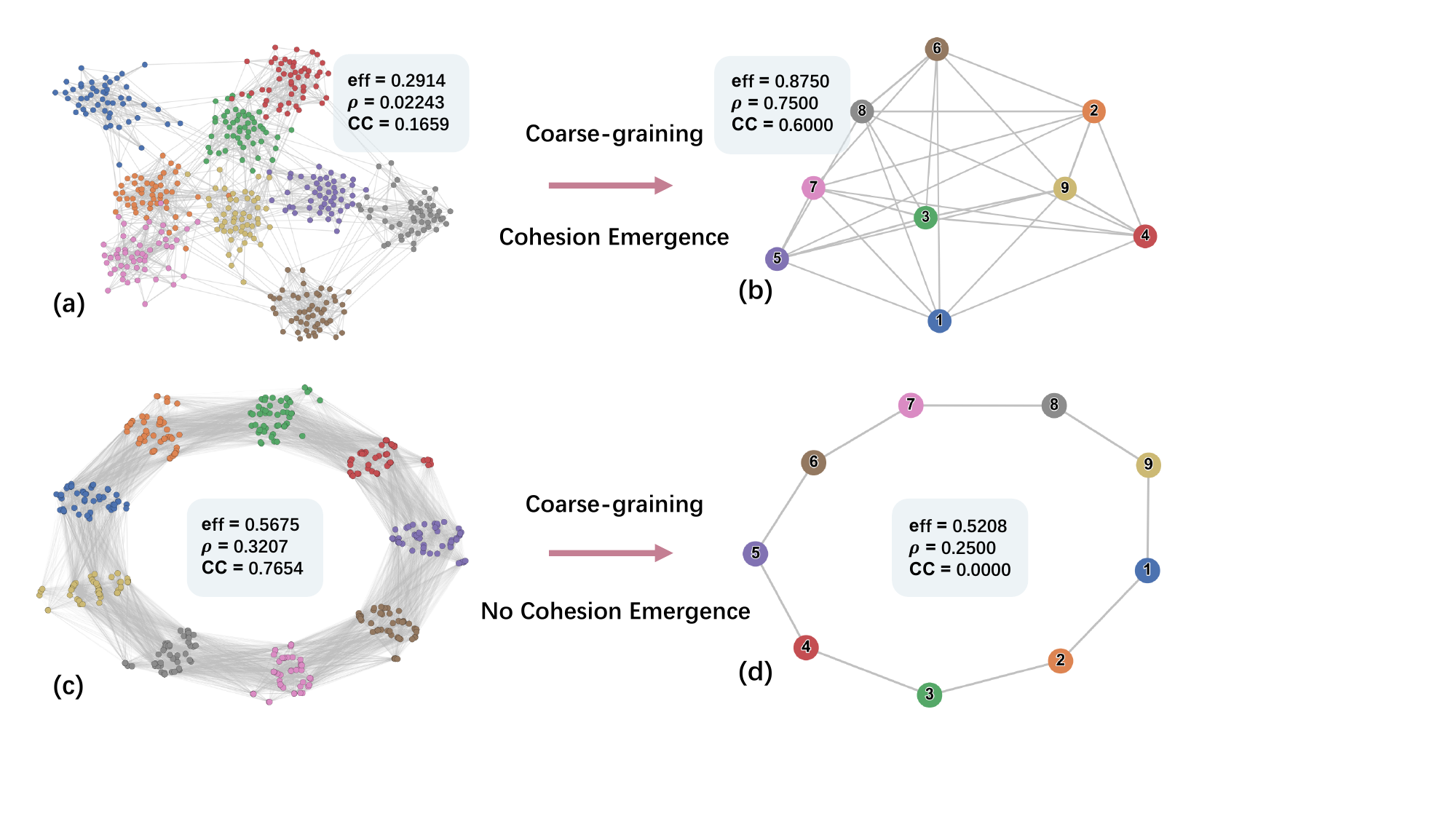}
    \caption{\textbf{Cohesion indicators under community-based coarse-graining.} Coarse-graining strategies combine each community of same-colored nodes in the microscopic networks $G(450)$ into macroscopic nodes, forming new macroscopic networks $G(9)$.
    The top row (Case 1, panels (a)--(b)) shows a weakly coupled modular
    network and its 9-community quotient graph. The bottom row (Case 2,
    panels (c)--(d)) shows a network of densely connected communities arranged
    in a ring and its 9-node ring quotient graph. In Case 1, global efficiency,
    edge density, and clustering increase under coarse-graining, so the system exhibits cohesion emergence. In Case 2,
    all three indicators decrease, and the system exhibits no cohesion emergence.}
    \label{fig:network}
\end{figure}

Interpret a network system \cite{latora2001efficient,fronczak2003mean,Liu2023} as an object composed of nodes, edges, and their
connectivity structure. We evaluate cohesion using three structural
indicators:
\begin{itemize}
    \item \textbf{Global efficiency $eff$}: measures the overall efficiency of
    information transfer in a network; shorter shortest paths correspond to
    higher global efficiency \cite{latora2001efficient}.
    \item \textbf{Edge density $\rho$}: the ratio of the number of edges to the
    maximum possible number of edges. Larger values indicate a denser network
    \cite{ronhovde2010local}.
    \item \textbf{Clustering coefficient $CC$}: measures local triangle
    closure by averaging, over nodes, the proportion of connections among
    each node's neighbors. Larger values indicate stronger local clustering
    \cite{fronczak2003mean}.
\end{itemize}
We take the multi-dimensional attribute to be
\begin{equation}
P=(eff,\rho,CC),
\end{equation}
and use the componentwise order, so that
\begin{equation}
P_1\prec P_2
\Longleftrightarrow
eff_1<eff_2,\quad \rho_1<\rho_2,\quad CC_1<CC_2.
\end{equation}
Thus, lifting relative to $P$ requires all three indicators to increase.

Let the research network be the system $\mathcal{S}$, with a network $G(450)$ with 450 nodes as microscopic
description category $\mathcal{C}_{\mu}$ and a network $G(9)$ with 9 nodes as macroscopic
community-level description category $\mathcal{C}_{M}$. A community
quotient functor $Q:\mathcal{C}_{\mu}\to\mathcal{C}_{M}$ maps nodes within
each community to a macro-level node and induces the connections between
communities in the quotient graph. We treat this community-based mapping as
an admissible coarse-graining; since it changes the granularity of the
description rather than merely relabeling nodes, it is also nontrivial.

In the first case, shown in Figure~\ref{fig:network} (a)--(b), the network
has nine densely connected communities, with relatively sparse connections
between communities. Its modular structure is therefore evident at the
microscopic level, while the quotient graph summarizes the connections among
the nine communities. The measured attributes are $P\bigl(G(450)\bigl)=(0.2914,0.0224,0.1659), P(G(9))=(0.8750,0.7500,0.6000)$.
%\begin{equation}
%\begin{aligned} 
%P\bigl(G(450)\bigr)
%    &= (0.2914,\,0.0224,\,0.1659),\\
%P\bigl(G(9)\bigr)
%    &= (0.8750,\,0.7500,\,0.6000).
%\end{aligned}
%\end{equation}
All three indicators increase, giving
\begin{equation}
P\bigl(G(450)\bigr)\prec P\bigl(G(9)\bigr).
\end{equation}
Accordingly, $\mathrm{Lift}_{P}(Q)$ holds in this case, and the coarse-graining
exhibits cohesion emergence relative to the selected indicators.

In the second case, shown in Figure~\ref{fig:network} (c)--(d), the nine
communities are internally dense and arranged in a ring, with neighboring
communities connected along the ring. Unlike the first case, its quotient
graph retains this ring arrangement. The measured attributes are $P\bigl(G(450)\bigr)=(0.5675,0.3207,0.7655), P\bigl(G(9)\bigr)=(0.5208,0.2500,0.0000)$.
%\begin{equation}
%\begin{aligned}
%P\bigl(G(450)\bigr)
%    &= (0.5675,\,0.3207,\,0.7655),\\
%P\bigl(G(9)\bigr)
%    &= (0.5208,\,0.2500,\,0.0000).
%\end{aligned}
%\end{equation}
Each indicator decreases under coarse-graining; hence
\begin{equation}
P\bigl(G(9)\bigr)\prec P\bigl(G(450)\bigr),
\end{equation}
and $\mathrm{Lift}_{P}(Q)$ does not hold. The two network cases therefore
show that cohesion emergence is not automatic: it is observed for the
weakly coupled modular network, but not for the ring-organized network, whose
three selected indicators all decline at the quotient level.

\section{Discussion}\label{sec:discussion}
The framework separates a structural question---how one description is obtained from another---from an evaluative question---whether a specified attribute improves.
Its contribution should therefore be assessed both by the comparisons this separation makes possible and by the scientific choices and mathematical constructions it leaves unresolved.

\subsection{Advantages of the Framework}
The main advantage is a common formulation that does not identify emergence with one particular mechanism or indicator.
Predictive efficiency in computational mechanics, effective information in causal emergence, and state-averaged approximate dynamical reversibility evaluate different aspects of a process \cite{Crutchfield1994,Hoel2013,Zhang2025}.
Their common structure is not equality of these quantities, but comparison of an attribute across descriptions connected by a specified coarse-graining.
The polygon and network examples show that this comparison can also be formulated without time evolution, probability, or intervention.
Their contrasting outcomes are important: the framework applies both when an attribute improves and when the same type of reduction fails to produce a lift.
These semantic choices belong to concrete instances rather than the basic definition, so disagreement between indicators need not be mistaken for a contradiction in the shared framework.

Separating admissibility, nontriviality, and lifting makes the grounds for an emergence claim explicit.
An allowed transformation may be only a relabeling, and a nontrivial transformation may fail to improve the attribute; conversely, a numerical gain does not establish that the transformation preserves the structure required by the application.
The conjunctive criterion distinguishes these possibilities instead of treating every reduction as emergence or every gain as sufficient evidence.
The updated examples give concrete instances of this distinction.
For polygons, the same 81-to-9 vertex-averaging scheme reduces both symmetry distances in one configuration but improves only one distance in the other; the latter does not satisfy the chosen joint improvement order.
For networks, the same 450-to-9 community reduction raises all three cohesion indicators in the weakly coupled modular case but lowers them in the ring-organized case.
Thus, even reductions to the same resolution need not produce the same sign of attribute change, and nontriviality alone does not establish lifting.
This also clarifies the relation to categorical accounts based on interconnection and generativity \cite{Adam2017,AdamDahleh2019}.
Whether observation preserves the composition of parts and whether a macroscopic description improves a selected attribute are different questions; neither answer generally determines the other.
The present framework complements such accounts by making the direction of improvement explicit rather than replacing their criteria.

The distinction is especially useful for dynamical models, where an increase in an indicator can coexist with an inaccurate macroscopic evolution rule.
Section~\ref{sec:dynamical-framework} separates the state map $\Phi$, the dynamical map $\Psi$, and their compatibility, so that closure concerns the validity of the macro dynamics while attribute lifting concerns the property being evaluated.
The discussion of effective-information maximization illustrates why these checks should not be conflated \cite{EberhardtLee2022,Yang2025}.
Evaluating a designated mechanism as an internal morphism also identifies exactly what is compared, without requiring the attribute assignment itself to be a functor.
This accommodates numerical indicators while retaining a structural correspondence between microscopic and macroscopic mechanisms.

Making observer dependence explicit improves the specification of a claim rather than reducing it to personal preference.
Different functors encode different choices about which distinctions to retain, and the attribute order states what counts as improvement.
Once these choices are fixed, the judgment is a mathematical comparison that another researcher can reproduce.
The static examples vary microscopic configurations under the same type of coarse-graining; they demonstrate dependence on system structure, not by themselves dependence on alternative observers of one fixed system.
The latter would require comparing different admissible functors for the same microscopic description and attribute.
Optimization in Section~\ref{sec:optimal-coarse-graining} adds a common selection rule: within a fixed admissible class, it asks which perspective best expresses the chosen attribute.
This can partially reduce dependence on arbitrary perspective selection, but only under the given attribute and constraints; it does not produce an observer-independent view.
The distinction between optimality and emergence further prevents the best available macro description from being declared emergent when it still fails to improve on the microscopic one.

Functoriality also gives a controlled way to discuss several description levels.
Identity and composition preservation make successive transformations structurally comparable, while the transitivity of the attribute order permits endpoint comparisons when the composite remains admissible and nontrivial.
This supports analysis of hierarchical descriptions without asserting that all coarse-graining chains improve monotonically.
The optional category $\mathrm{Desc}(\mathcal{S})$ organizes such alternatives and chains, but a single emergence claim needs only the relevant descriptions and functor.
Keeping these levels separate avoids treating a mathematical organization of descriptions as a further physical component of the system.

\subsection{Limitations of the Framework}
The breadth of the definition comes at the cost of substantial application-specific input.
Category theory alone cannot determine which descriptions represent the same system, which distinctions may be discarded, or which property is scientifically important.
Without independently justified choices of $P$, its order, and the admissible class, the criterion risks becoming a restatement that a preferred description scores better by a preferred measure.
Attributes and constraints should therefore be motivated before selecting a favorable coarse-graining, and competing choices should be examined when the scientific interpretation depends on them.
The resulting judgment does not establish strong emergence, ontological novelty, or irreducibility in the philosophical sense \cite{Bedau1997,Chalmers2006}.

Cross-level comparability is a substantive requirement, not a consequence of using the same symbol $P$.
The evaluations must refer to the same property with compatible units, normalizations, and observational conventions.
The new network comparison keeps the microscopic and macroscopic node counts equal across the two cases, so their opposite changes cannot be explained by the reduction in node count alone.
Nevertheless, contracting a community removes its internal distinctions and changes the meaning of path length, edge density, and triangle closure at the quotient level.
An increase in these structural statistics is therefore not an independent demonstration of enhanced physical communication.
Likewise, the matched 81-to-9 polygon comparisons reduce a resolution mismatch between cases, but the reference symmetry classes still depend on vertex count and require consistent distance normalization.
These matched sizes strengthen the comparisons without replacing suitable null models, checks of weighting conventions, or tests of robustness across configurations.

A further limitation concerns the construction of the categories and functors themselves.
Calling vertex averaging or a community quotient a coarse-graining does not yet specify its action on every relevant morphism or prove preservation of composition.
A fully instantiated categorical claim needs an explicit representation in which these properties can be checked.
The illustrative examples establish attribute comparisons, but do not supply a complete construction of all the corresponding description categories and morphism actions.
Similarly, a state map and a fitted macro kernel do not automatically extend to a single functor.
The strict special case in Section~\ref{sec:dynamical-framework} must not be substituted for that general modeling situation, and an approximate closure bound is not itself a proof of exact functoriality.
Developing concrete constructions and clarifying how approximate models relate to exact categorical witnesses remain important tasks.

Nontriviality also remains relative to the structures a description is intended to preserve.
Excluding categorical equivalences rules out pure descriptive changes, but failure to be an equivalence is not a sufficient account of meaningful macroscopic organization.
A transformation can destroy relevant information without creating a useful description, and a collapse can improve a selected statistic while violating the scientific purpose of the model.
Admissibility must exclude such cases through justified preservation or accuracy constraints; the abstract framework cannot supply those constraints universally.
When additional geometry, probability, or observational structure matters, the equivalence convention and the definition of nontriviality must take that structure into account.
The improvement order is also consequential: the second polygon case fails the joint criterion even though one symmetry component improves.
Selecting that component alone would give a different lifting judgment, whereas the network ring case lowers all three selected indicators.
These are distinct ways of failing the chosen joint criterion; neither failure implies that no other attribute or admissible coarse-graining could witness emergence.

Finally, formal selection of an optimum does not resolve computational or empirical uncertainty.
Even finite partition searches may be too large for exhaustive evaluation, while an unrestricted infinite candidate class need not attain its supremum.
Several functors can share an optimal target, and multiple attributes may yield incomparable Pareto-optimal descriptions rather than a unique best perspective.
These alternatives preserve observer dependence despite the common optimization rule.
For models inferred from data, estimation error can also change the apparent attribute gain or closure defect, and maximizing over many candidates can favor a fit that does not generalize.
Uncertainty estimates, independent validation data, and sensitivity to constraints are therefore needed for empirical claims.
The framework currently provides neither a general search algorithm nor such statistical guarantees; domain-specific constructions and validation are needed to turn its formal comparisons into reliable scientific conclusions.

\section{Conclusion}
This paper proposes a category-theoretic framework for generalized emergence: a system admits multiple structured descriptions, each represented by a category; coarse-graining from micro to macro is a functor $Q$ between description categories; and generalized emergence is an attribute lift along an admissible, nontrivial $Q$, with $\mathrm{Emergence}_P(Q)\iff \mathrm{Adm}(Q)\wedge\mathrm{Nontrivial}(Q)\wedge\mathrm{Lift}_P(Q)$, where attributes evaluate descriptions through any part of their data, including designated mechanisms.
Because $Q$ embodies an observational perspective, the judgment is observer-dependent; seeking an optimal coarse-graining can partially reduce this dependence, but only relative to the chosen attribute, its improvement order, and the constraints defining the admissible class.
The framework presupposes no dynamical or probabilistic semantics: it accommodates static structural instances as well as dynamical ones, in which distributions and Markov kernels are morphisms, state coarse-graining and dynamical coarse-graining are constructed separately and linked by strict or approximate closure conditions, and attributes such as effective information and approximate dynamical reversibility evaluate the designated mechanisms.
What category theory provides is not a ``super-indicator'' that replaces concrete metrics, but a meta-theoretical language that clearly distinguishes shared structure from concrete differences.
Turning this language into a computable, comparable, and verifiable scientific theory requires constructing concrete admissible coarse-graining classes and studying the precise relation between nontriviality and categorical equivalence, the existence and uniqueness of attribute-optimal coarse-grainings, and approximate coarse-graining and approximate closure theories in metric categories, enriched categories, or other structures with errors.

\bibliographystyle{ieeetr}
\bibliography{jiaoda11}

\end{document}